\documentclass[reqno]{amsart}
\usepackage{graphicx}
\usepackage{textcomp}

\begin{document}

\title[Major feedback]{Major feedback factors and effects of the cloud cover and the relative humidity on the climate}%
\author{J. Kauppinen and P. Malmi}%
\address{Department of Physics and Astronomy, University of Turku}%
\email{jyrkau@utu.fi}


\date{\today}%

\begin{abstract}
In this paper we derive a new formula for the global temperature change and major feedback portions in the climate response. In our earlier paper \cite{Kau2011} we calculated from the experimental values the sensitivity about 0.058~K/(W/m$^2$). This means the negative feedback which reduced the sensitivity by factor 2.13. In this paper we explain and derive the major portions in the feedback coefficient using the observed energy budget at the top of the climate and on the surface of the earth. The results also support strongly our earlier results of the low climate sensitivity ($\Delta T_{\rm 2CO_2}\approx 0.24^{\circ}$C). The major portions in the negative feedback coefficient in shortwave insolation are roughly clouds $63\%$, evaporation cooling $28\%$, and water vapour $9\%$. The new sensitivity is 0.0605~K/(W/m$^2$) which is reduced by factor 2.00. The changes in cloud cover or in the relative humidity explain almost all the global temperature changes. The result is confirmed with experimental observations \cite{Dou2004Jan}. On the other hand, the sun and the change in the vegetation are probably controlling most of the changes in cloudiness and humidity.
\end{abstract}
\maketitle

\section{Introduction}

\begin{figure}[h]
  \centering
  \includegraphics[width=\columnwidth]{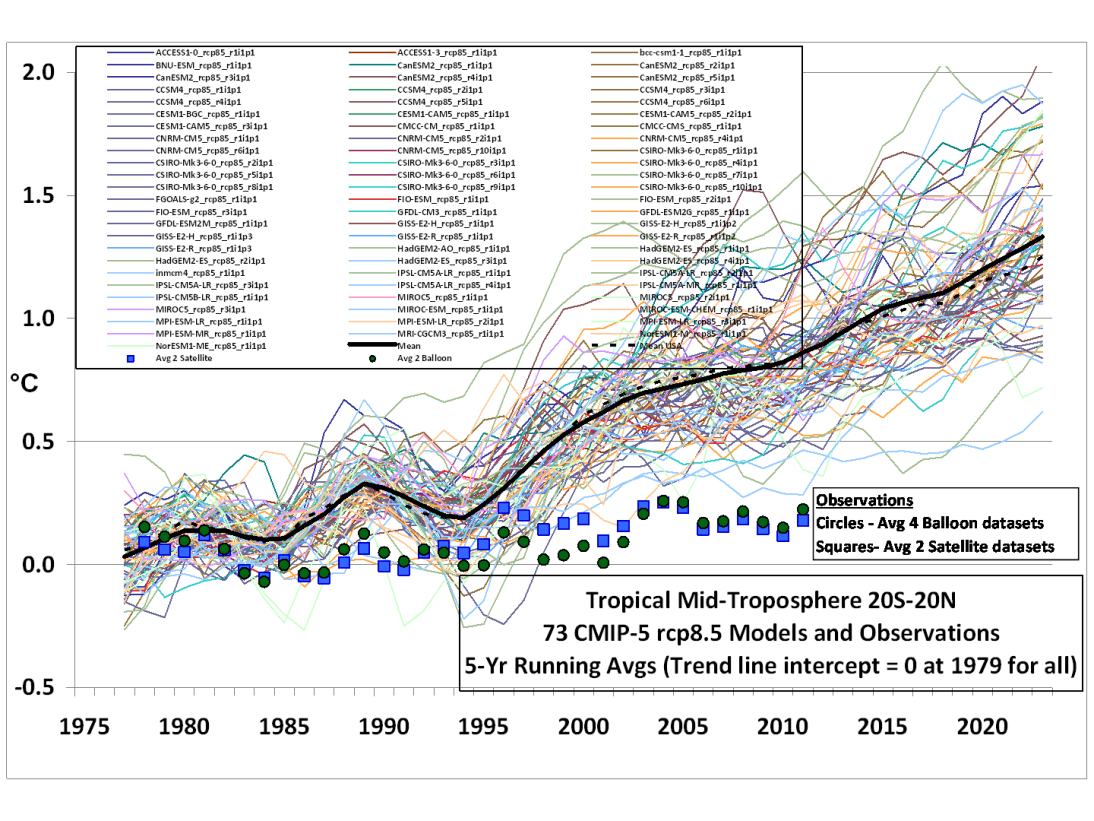}
  \caption{73 models vs observed}
  \label{CMIP5}
\end{figure}

The main goal of this study is to calculate the change of the global mean temperature of the earth's climate due to a forcing like greenhouse gases and also without forcing. The reported temperature changes $\Delta T_{\rm 2CO_2}$ of the climate due to the doubling of the CO$_2$ concentration are still within a very large uncertainty range. According to the Intergovernmental Panel on Climate Change (IPCC) the change of the global mean temperature $\Delta T_{\rm 2CO_2}$ is likely between 2 and 4.5~K, most likely 3.2~K \cite{IPCC}. Hansen et al. \cite{Han1984} have reported $\Delta T_{\rm 2CO_2}$ between 2 and 5~K, assuming that the present temperature change results from the increased concentration of greenhouse gases. The major reason to the uncertainty is that the sensitivity $R=dT/dQ$ of the climate is not very well known. The sensitivity gives us the surface temperature change $\Delta T = R\Delta Q$, where $\Delta Q$ (W/m$^2$) is the radiative forcing. Values of IPCC and Hansen imply that there is a positive net feedback in the climate system. Climate sensitivity estimated by IPCC is based mainly on theoretical circulation models. Figure~\ref{CMIP5} shows an example how well these models can predict the global temperature. However, there are papers by Douglass et al. \cite{Dou2004Jan,Dou2004Mar} and Idso \cite{Ids}, where much smaller sensitivities are presented. These results are obtained e.g. from the annual solar irradiance cycle. Also Lindzen \cite{Lin2007}, Lindzen and Choi \cite{Lin2009}, Spencer and Braswell \cite{Spe2008,Spe2010} have reported a net negative feedback.

As shown in Fig. \ref{CMIP5} the temperature between years 1998 and 2010 has been almost constant even though the models predict an average increase of more than 0.6~K. The newest observed temperature data between 2010 and 2015 are also at the same level. This is why we conclude that some new approach to the climate model has to be found. This is already done in our first paper \cite{Kau2011}, where we derived the sensitivity and the response time of climate based on experimentally measured data of the climate. Two completely different methods using different observations lead to the sensitivity of about $0.058~{\rm K/(W/m^2)}$. One of these methods gave also the response time of the climate about 1.3 months. The small sensitivity implies that the net feedback is negative.

\begin{figure}
  \centering
  \includegraphics[width=\columnwidth]{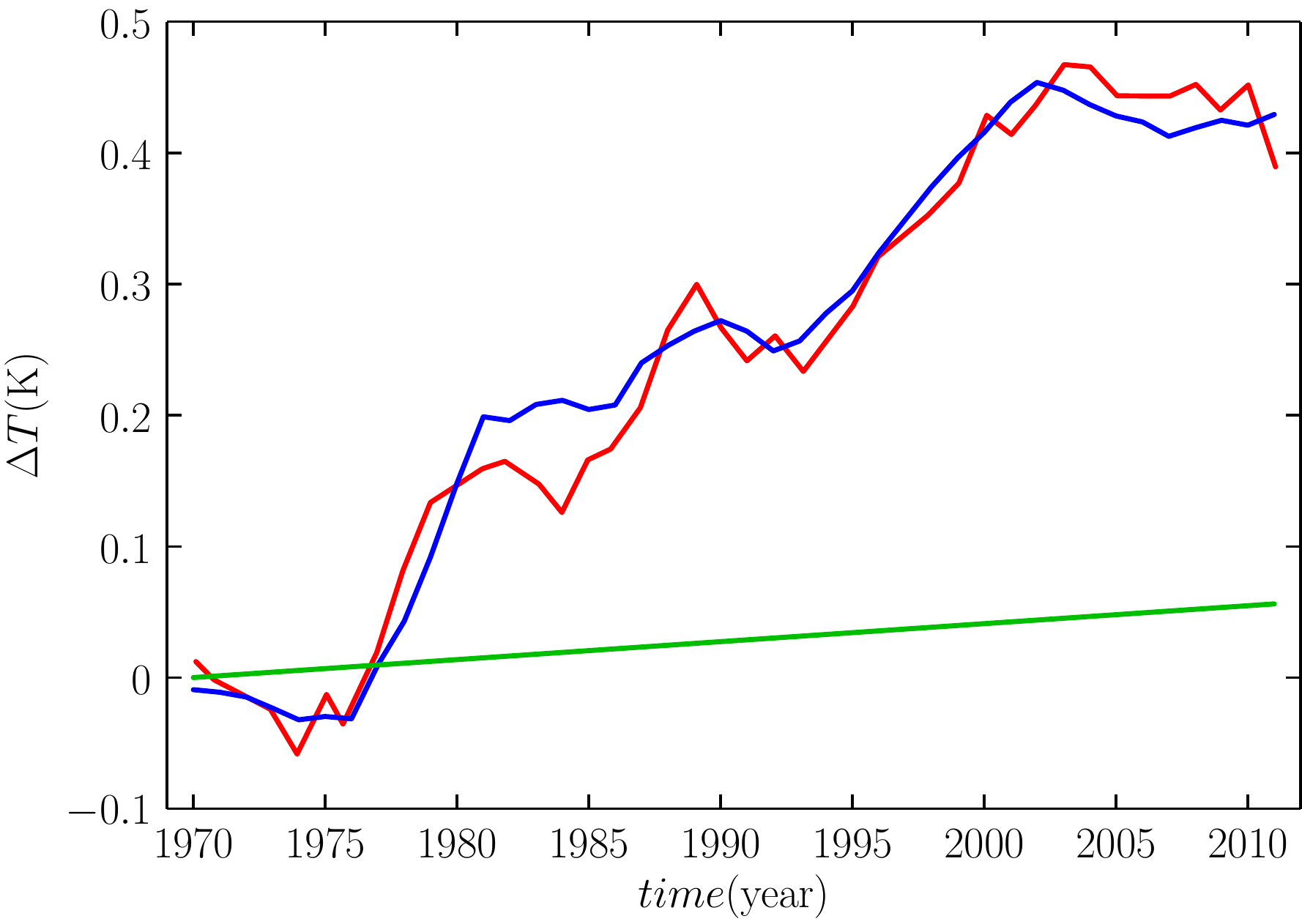}\\
  \caption{Observed global mean temperature anomaly (red), calculated anomaly (blue) and contribution of CO$_2$ (green)}
  \label{Fig8}
\end{figure}

Our second paper \cite{Kau2014} gives an example of the application of the first paper \cite{Kau2011} to the real climate. This example is shown in Fig. \ref{Fig8}. Compare the figures \ref{CMIP5} and \ref{Fig8}. However, in the previous papers we were able to derive only the net feedback. In this paper we will derive the sensitivity and the major components of the feedback using observed values of the climate different from the ones in the previous paper. In addition, we will study the role of clouds and the relative humidity in the atmosphere and we will make some remarks of the general circulation models.

\section{Basic definitions}

\begin{figure}
  \centering
  \includegraphics[width=\columnwidth]{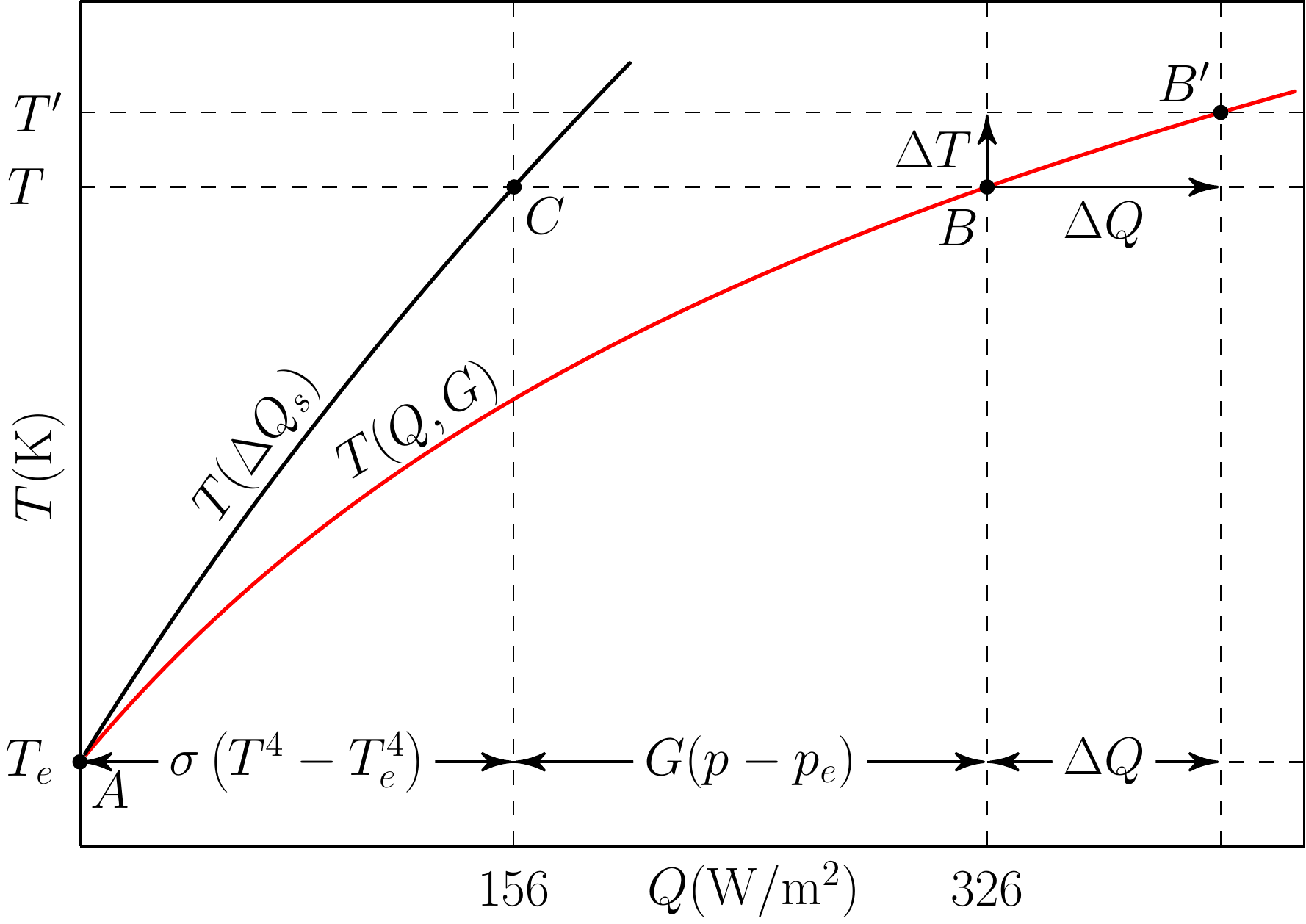}\\
  \caption{$Q_s=\sigma T^4,\Delta Q_s=\sigma(T^4-T_e^4)$, $\sigma$ is the Stefan-Boltzmann constant, $T=289$~K, and $T_e=255$~K.
  The total forcing is $Q=\sigma(T^4-T_e^4)+G(p-p_e)$ and $\Delta T=T'-T=R\Delta Q$.}
  \label{TQG}
\end{figure}

In our previous paper \cite{Kau2011} we used the model, where the total forcing $Q(T)$ i.e. the net absorption in the longwave emission is
\begin{equation}\label{Q(T)}
  Q(T)=\Delta Q_s(T)+G[p(T)-p(T_e)]=\epsilon Q_s(T),
\end{equation}
where $\Delta Q_s(T)=\sigma(T^4-T_e^4)$ is the difference between the longwave emissions at the present temperature $T$ and the reference temperature $T_e$. Similarly ${p(T)-p(T_e)}$ is the difference between the water vapour saturation pressures and $G$ is the proportionality coefficient.

In figure 3 the red curve goes through the points A and B. A is the point, where $T_e$ is 255~K i.e. the temperature with zero total forcing and $B$ is the present point with $T=289$~K. We define the climate sensitivity as
\begin{equation}\label{R}
  R=\frac{dT}{dQ},
\end{equation}
which gives us the temperature change
\begin{equation}\label{DeltaT}
  \Delta T=\frac{dT}{dQ}\Delta Q
\end{equation}
due to the forcing $\Delta Q$. The sensitivity $R$ is the derivative of the curve $T(Q,G)$ at point $B$. In addition, we define
\begin{equation}\label{R_0}
  R_0=\frac{dT}{dQ_s}=\frac{T}{4\sigma T^4} \approx 0.182~{\rm K/(W/m^2)},
\end{equation}
which is the climate sensitivity without the negative feedback, i.e. the derivative of $T(\Delta Q_s)$ at the point $C$. The differentiation of Eq. \ref{Q(T)} with temperature gives
\begin{equation}\label{dQ/dT}
  \frac{dQ}{dT}=\frac{dQ_s}{dT}+G\frac{dp}{dT}
\end{equation}
or
\begin{equation}\label{1/R}
  \frac{1}{R}=\frac{1}{R_0}+G\frac{dp}{dT}.
\end{equation}
The above equation gives us the sensitivity $R$ as follows
\begin{equation}\label{R2}
  R=\frac{R_0}{1+R_0G\frac{dp}{dT}},
\end{equation}
where the feedback coefficient is
\begin{equation}\label{f(T)}
  f(T)=-G\frac{dp}{dT}.
\end{equation}
Now Eq. \ref{Q(T)} can be rewritten in a differential form
\begin{equation}\label{dQ}
  dQ=dQ_s-f(T)dT.
\end{equation}
So $f(T)$ is negative and explains mainly the negative feedback in the shortwave insolation. The possible positive feedback takes place in longwave absorption and we calculate it separately for the greenhouse gases as we did in our previous papers, too. In the present atmosphere the measured total forcing 326~W/m$^2$ \cite{Kie1997} includes all the possible forcings like water vapour, clouds, and all the greenhouse gases and so on. The positive feedback is also included in the total forcing. If we like to derive the sensitivity for greenhouse gases like CO$_2$ we have to take into account the positive feedbacks of water vapour and clouds in forcing. These feedbacks increase the forcing via $\Delta Q\approx\Delta Q_0+1~{\rm W/(m^2K)}\Delta T$, where the feedback coefficient 1~W/(m$^2$K) is the sum of water vapour and cloud contributions. This can be easily taken into account in $R$ by substituting $Gdp/dT-1$~W/m$^2$K for $Gdp/dT$.

\begin{figure}
  \centering
  \includegraphics[width=\columnwidth]{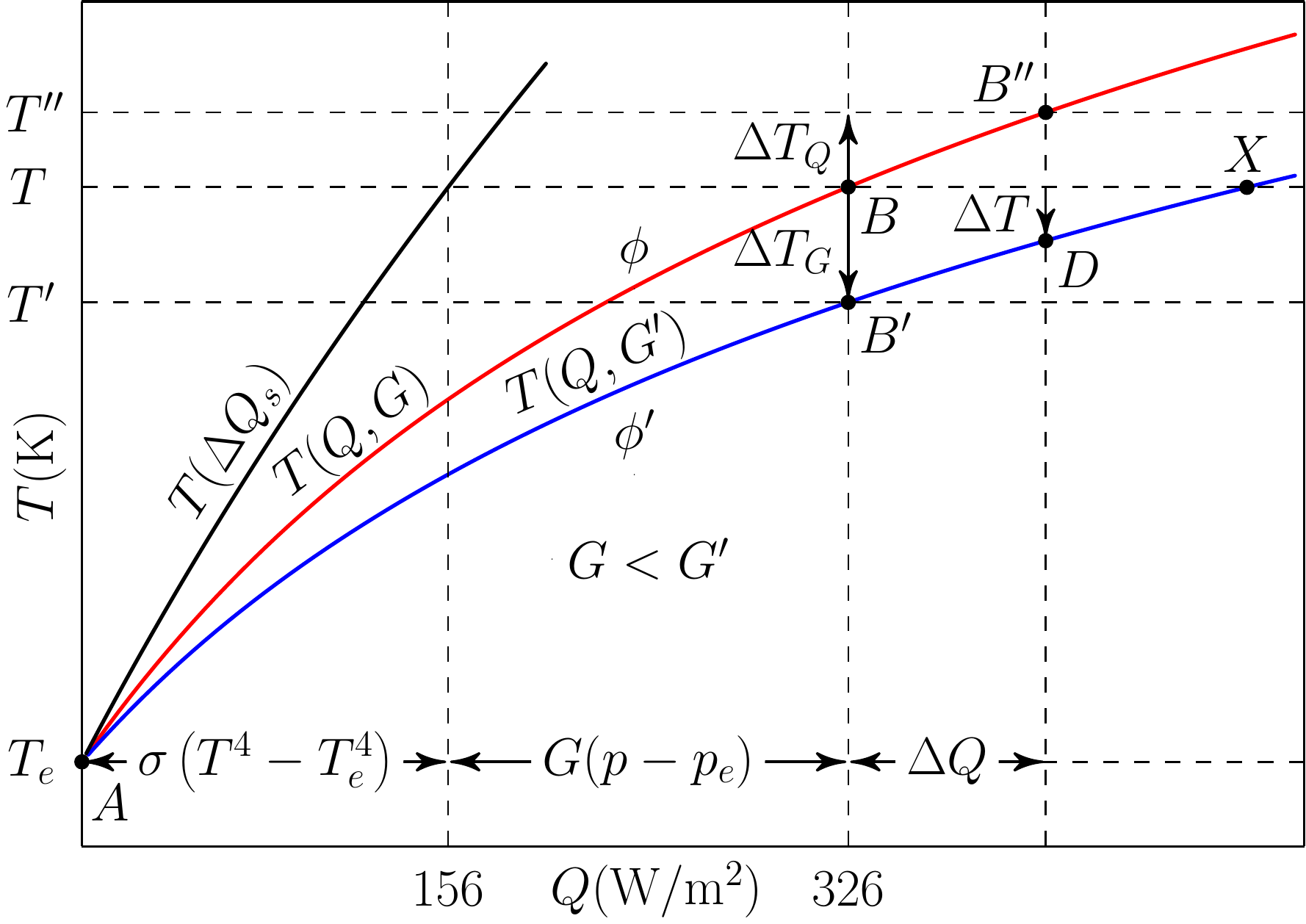}\\
  \caption{$Q_s=\sigma T^4,\Delta Q_s=\sigma(T^4-T_e^4)$, where $\sigma$ is the Stefan-Boltzmann constant, $T=289$~K, and $T_e=255$~K.
  The total forcing is $Q=\sigma(T^4-T_e^4)+G(p-p_e)$.
  The total temperature change in the process $BB''D$ or $BB'D$ is $\Delta T=\Delta T_Q+\Delta T_G$.
  Note that $\Delta T_Q$ is positive and $\Delta T_G$ is negative in this figure.
  In the process $BB'$ $\Delta Q=0$, but $G$ and $R$ are changing.}\label{TQGG}
  \label{TOGG}
\end{figure}

In Fig. \ref{TQGG} the relative humidities are the constants $\phi$ and $\phi'$ along the curves $T(Q,G)$ and $T(Q,G')$, respectively, because $G=constant\times\phi$. The sensitivities $R$ or the derivative values at the points $B$ and $B''$ are very close to each other but a few per cent larger than the derivative values at the points $B'$, $D$, and $X$ (on the blue curve), which are in turn almost equal. According to this figure the total temperature change is
\begin{equation}\label{DeltaT3}
  \Delta T=R\Delta Q+R\Delta G(p-p_e)=\Delta T_Q+\Delta T_G.
\end{equation}
The derivation of the second term $\Delta T_G= R\Delta G(p-p_e)=-RG(p-p_e)d\phi/\phi$ is given in our second paper \cite{Kau2014}. This is a very interesting result. The red curve $T(Q,G)$ goes through the points $A$ and $B$, if $G=102$~W/(m$^2$kPa). In this way calculated $G$ gives us the $R=0.0577$~K/(W/m$^2$), according to Eq. (\ref{R2}). See more details in our first paper \cite{Kau2011}. A doubling of the CO$_2$ concentration in the atmosphere has been estimated by IPCC to cause an additional forcing $\Delta Q=3.78$~W/m$^2$ which would lead to an increase of the average surface temperature by $(0.0577\cdot 3.78)~{\rm K}=0.22~{\rm K}$.

\begin{figure}
  \centering
  \includegraphics[width=\linewidth]{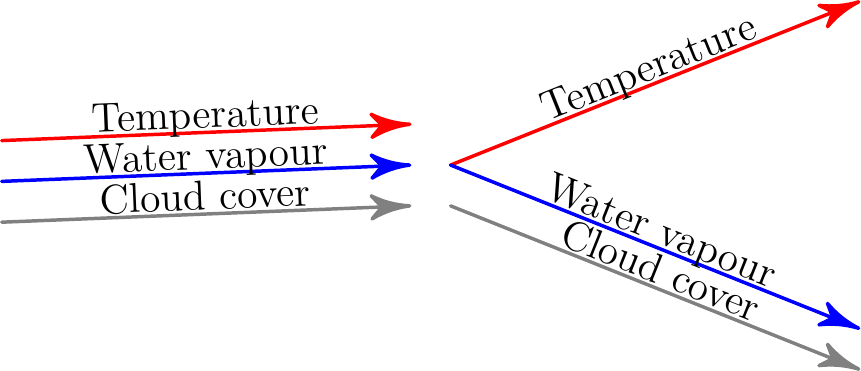}\\
  \caption{Left) Greenhouse effect, the first term in Eq.~\ref{DeltaT3}.
  Right) Hydrological cycle, the second term in Eq.~\ref{DeltaT3}.}
\label{pilvet}
\end{figure}

The above equations state that a very small change of $R$ or $G$ results in a remarkable temperature change. So, it turns out that the last term is dominating in the real climate due to the change of $G$ or $R$. All the observed temperature changes are originating from the second term. The first term $R\Delta Q$ (the order of 0.1~K) is so small that it is very difficult to measure.

Figure 5. shows the contribution of the forcing $\Delta Q$, i.e. the greenhouse effect, or the first term in Eq. (\ref{DeltaT3}). In this case both temperature and clouds (or relative humidity) change to the same direction, but very little ($\Delta T<0.1$~K). However, in the second term $-RG(p -p_e)\,d\phi/\phi$, the cloud amount and temperature change in opposite directions and are typically order of 1~K.

\section{Major feedback coefficients and the sensitivity of the climate using all the global energy flows}

\begin{figure}
  \centering
  \includegraphics[width=\columnwidth]{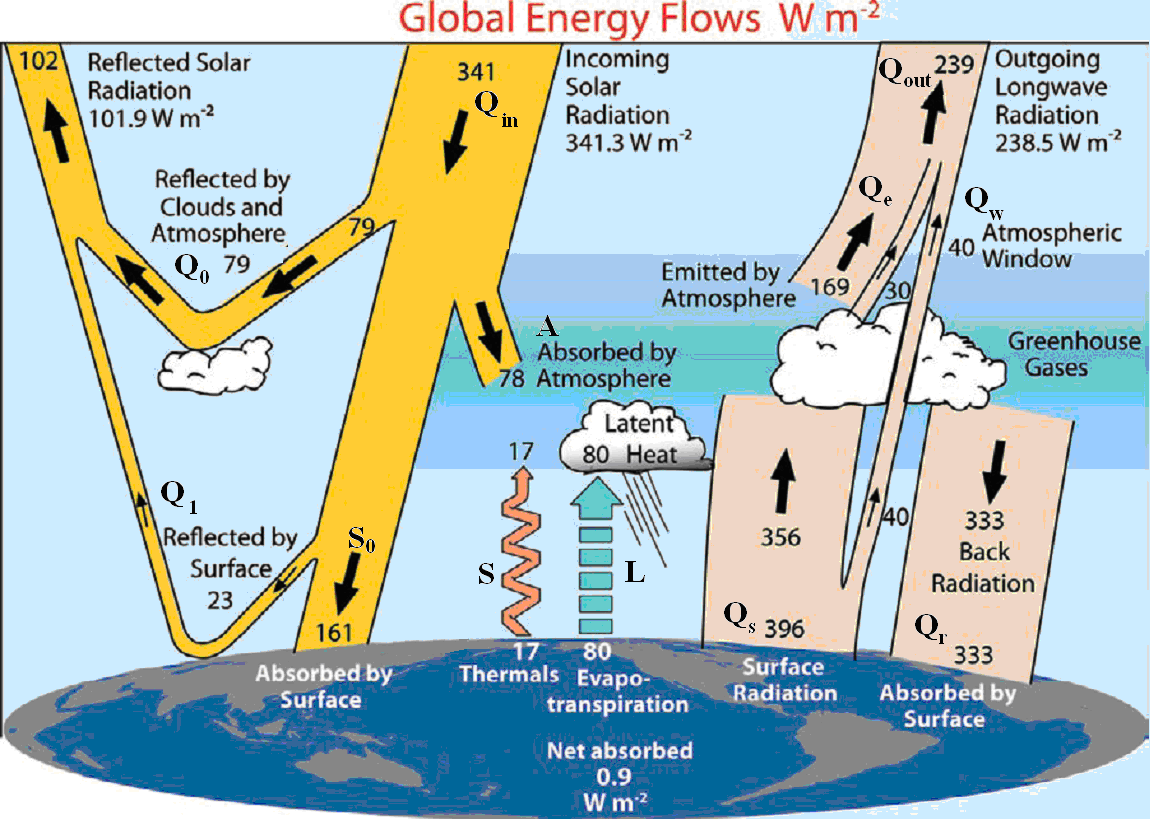}\\
  \caption{Global energy budget. All numerical values in the picture are in units W/m$^2$.}
  \label{GEF}
\end{figure}

We will derive the major feedback coefficients and the sensitivity of the climate using merely an experimental global energy budget of the climate. The energy budget is presented in Fig. \ref{GEF} and the experimental values are from Fig. 1 in Ref. \cite{Tre}. The difference between this treatment and the earlier papers is the fact that in this presentation we don’t use the point $A$ shown in Fig. \ref{TQG} but we derive all the quantities around the point $B$ at the present climate conditions using the global energy budget shown in Fig. \ref{GEF}.

At a balance we have the condition at the top of the climate
\begin{equation}\label{topbalance}
  Q_{in}=Q_0+Q_1+Q_e+Q_s(1-\epsilon),
\end{equation}
where $Q_s(1-\epsilon)$ is the transmitted flux. At the surface of the earth we have
\begin{equation}\label{surfacebalance}
  S_0+Q_r=S+L+Q_s.
\end{equation}
It is also possible to write the balance equation in the atmosphere between the top of the climate and the surface, but it is linearly dependent on Eq. \ref{topbalance}. For example, we have
\[ 
  A+S+L+Q_s=Q_e+Q_s(1-\epsilon)+Q_r
\] 
If we differentiate the Eqs. (\ref{topbalance}) and (\ref{surfacebalance}) with respect to the temperature T, we have three  unknown derivatives $dQ_r/dT$, $dQ_e/dT$ and $dQ_s/dT$. Because we have only two linearly independent equations, we cannot solve all the three derivatives. However, we have $Q_e=rQ_r$, where $Q_e$ is the IR emission up and $Q_r$ the emission down from the atmosphere and $r$ is constant 0.5075 in a small change ($dr/rdT < 3 \cdot 10^{-3}$~1/K). In addition, $Q_r$ and $Q_e$ depend on $\epsilon$. Our aim is to solve $dQ_s/dT$, because $Q_s=\sigma T^4$ is in the relation to the surface temperature $T$, if we assume the emissivity to be one.

Next we solve the problem using equations (\ref{topbalance}) and (\ref{surfacebalance}). The forcing is the sum of the changes including all the terms, which change due to the change of absorption in longwave emission. For example at the top of the climate differentiating Eq. (\ref{topbalance}) the forcing is the sum of $\Delta Q_e$ and $\Delta (\epsilon Q_s )$ but according to Eq. (\ref{surfacebalance}) at the surface the forcing is $\Delta Q_r$, respectively. So the forcing depends on the equation we use. Further the sensitivity $R$ depends also on the used forcing or the selected coordinate axis $Q$. However the temperature change must be the same in all cases, or
\begin{equation}\label{DeltaT2}
  \Delta T_Q=\frac{\partial T}{\partial Q}\Delta Q=R\Delta Q=\frac{\partial T}{\partial Q'}\Delta Q'=R'\Delta Q'
\end{equation}
if $Q'=constant\times Q$.

The dependencies of $Q_e$ and $Q_r$ on $T$ and $\epsilon$ are complicated to derive, because we have to use the Schwartzschild equations \cite{Har}, which give us $Q_e$ and $Q_r$ after very hard computation. In other words the derivatives $dQ_e/dT$ and $dQ_r/dT$ are not easy to calculate. That is why we eliminate them from Eqs. (\ref{topbalance}) and (\ref{surfacebalance}) and we have left only their ratio $r$, which does not depend much on the temperature $T$ in small changes. Solving $Q_r$ from Eq. (\ref{surfacebalance}) using $Q_e=rQ_r,\epsilon Q_s=Q$ and $S_0=Q_{in}-Q_0-A-Q_1$ in Eq. (\ref{topbalance}) we can solve $Q_s$ as follows:
\begin{equation}\label{Q_s}
  Q_s=Q_{in}-Q_0-Q_1-\frac{r}{r+1}(S+L+A)+\frac{Q}{r+1}.
\end{equation}
Differentiation with respect to $T$ gives
\begin{equation}\label{dQ_s/dT}
  \frac{dQ_s}{dT}=\frac{dQ_{in}}{dT}-\frac{dQ_0}{dT}-\frac{dQ_1}{dT}-%
  \frac{r}{r+1}\left(\frac{dS}{dT}+\frac{dL}{dT}+\frac{dA}{dT}\right)+\frac{1}{r+1}\frac{dQ}{dT}
\end{equation}
or
\[ 
  \frac{1}{R_0}=-Z+\frac{1}{r+1}\frac{1}{R}
\] 
and finally
\begin{equation}\label{R3}
  R=\frac{R_0/(r+1)}{1+ZR_0}=\frac{R_{00}}{1+(r+1)ZR_{00}},
\end{equation}
where $R_{00}=R_0/(r+1)$ is the sensitivity with zero feedback $Z$ and
\begin{equation}\label{Z}
  Z=-\frac{dQ_{in}}{dT}+\frac{dQ_0}{dT}+\frac{dQ_1}{dT}+\frac{r}{r+1}\left(\frac{dS}{dT}+\frac{dL}{dT}+\frac{dA}{dT}\right).
\end{equation}
According to Fig. \ref{GEF}, in shortwave insolation, $(Q_0+Q_1)/Q_{in}$ is the planetary albedo and $Q_1/(Q_{in}-Q_0-A)$ is the surface albedo. Because $Q_0, Q_1$, and $A$ depend on each other, we have to continue taking these relations into account. According to Fig. \ref{GEF}
\begin{equation}\label{Q_0}
  Q_0=r_0Q_{in}\quad{\rm or}\quad \frac{dQ_0}{dT}=\frac{dr_0}{dT}Q_{in}=\frac{dr_0}{r_0dT}Q_0,
\end{equation}
where $r_0= 0.232$ and $dQ_{in}/dT=0$. The absorption in shortwave insolation is
\begin{equation}\label{A}
  A=\epsilon_0(Q_{in}-Q_0),
\end{equation}
where $\epsilon_0=0.298$ and
\begin{equation}\label{dA/dT}
  \frac{dA}{dT}=\frac{d\epsilon_0}{\epsilon_0 dT}A-\epsilon_0\frac{dQ_0}{dT}.
\end{equation}
The reflection from the surface of the earth is
\begin{equation}\label{Q_1}
  Q_1=\alpha_s(Q_{in}-Q_0-A)
\end{equation}
or
\begin{equation}\label{dQ_1/dT}
  \frac{dQ_1}{dT}=-\alpha_s\left(\frac{dQ_0}{dT}+\frac{dA}{dT}\right),
\end{equation}
where $\alpha_s=23/184=0.125$ is the surface albedo and $dQ_{in}/dT=0$ or no solar forcing. All the negative feedbacks ($-Gdp/dT$) take place in shortwave insolation due to increasing water content of the atmosphere with increasing temperature. The proportionality coefficient $G$ is proportional to the relative humidity $\phi$, which is $p_t/p$, where $p_t$ is the absolute partial pressure of water vapour. The relative change of $\phi$ is given by
\begin{equation}\label{dphi/phi}
  \frac{d\phi}{\phi}=\frac{dp_t}{p_t}-\frac{dp}{p}.
\end{equation}

If the relative humidity is constant, then $dp_t/p_t=dp/p$. In order to calculate all the above derivatives, we assume that the relative humidity is constant which means that $G$ is constant, too. Note that in Fig. \ref{TQG} the relative humidity $\phi$ is constant along the red curve $T(Q,G)$. Thus, here we assume that all the derivatives are proportional to the change of water vapour and clouds, and we have
\begin{equation}\label{dp/pdT}
  \frac{dr_0}{r_0dT}=\frac{d\epsilon_0}{\epsilon_0 dT}=\frac{dp}{pdT}=\frac{dp_t}{p_t dT}.
\end{equation}
Later we see that the key process is the mass flow through the atmosphere. In addition $dL/LdT=dp/pdT$, too. All these terms of shortwave insolation include $dp/pdT=0.0641/{\rm K}$ and are constant over quite a large temperature range. Note that $p$ and $dp/dT$ both have an exponential shape \cite{Kau2011}. Now we are able to write down the final derivatives in $Z$. The result is
\begin{eqnarray}\label{Z2}
    Z&=&\frac{dp}{pdT}\left[\left(1-\alpha_s+\alpha_s\epsilon_0-\frac{r\epsilon_0}{r+1}\right)Q_0
    +\left(\frac{r}{r+1}-\alpha_s\right)A+\frac{r}{r+1}(S+L)\right]\nonumber\\
     &=& 6.253~{\rm (W/m^2)/K},
\end{eqnarray}
where $\alpha_s=0.125$, $\epsilon_0=0.298$, $r=0.5075$, $dp/pdT=0.0641$~1/K, $S=17$~W/m$^2$, and $L=80$~W/m$^2$. In the above equation we use $Q_0\approx 68.2$~ W/m$^2$ and $A=45$~W/m$^2$. In Fig. \ref{GEF} $Q_0$ is 79~W/m$^2$ but we have subtracted the portion scattered by air. Also in $A=78$~W/m$^2$ only the absorption 45~W/m$^2$ of clouds and water vapour has been used \cite{Kie1997}. Now, according to Eq. (\ref{R3})
\begin{equation}\label{R3}
  R=\frac{R_{00}}{1+(r+1)ZR_{00}}=0.0565~{\rm K/(W/m^2)}
\end{equation}
where $R_{00}=R_0/(r+1)=0.1207$~K/(W/m$^2$) is now the sensitivity without feedback. The feedback coefficient
\begin{equation}\label{Gdp/dT}
  -G\frac{dp}{dT}=-(r+1)Z,
\end{equation}
which gives $G=82$~W/(m$^2$ kPa). The other way to calculate G is the relation
\begin{equation}\label{Gdp/dT2}
  G\frac{dp}{dT}=\frac{1}{R}-\frac{1}{R_{00}}.
\end{equation}
As pointed out before we can say that $G, R, R_{00}$ and $\Delta Q$ depend on the used balance equation like Eq. (\ref{Q_s}) and on the defined coordinate axes $Q$. In the above presentation $dQ=d(\epsilon Q_s)=d\epsilon Q_s+\epsilon dQ_s$, which is the same as in our earlier paper. In addition to solar forcing $\Delta Q_{in}$ we can define from Eq. (\ref{Q_s}) a few more forcings like $Q_s d\epsilon, Q_s d\epsilon/(r+1)$, and $d(\epsilon Q_s)/(r+1)$. Note that all the forcings give the same $\Delta T$.

\section{The Temperature Change without the Change in Greenhouse Gases}

In our earlier papers \cite{Kau2011,Kau2014} we used the total forcing given in Eq. (\ref{Q(T)}). Now we will use a little different presentation given by \begin{equation}\label{Q2(T)}
  Q(T)=(r+1)\Delta Q_s(T)+Gp=(r+1)\sigma(T^4-T_e^4)+Gp=\epsilon Q_s(T),
\end{equation}
because the derivation of this equation gives Eq. (\ref{Gdp/dT2}). Note that in earlier papers we did not use the global energy flow budget shown in Fig. \ref{GEF}, which gives an extra number $r$.

\begin{figure}
  \centering
  \includegraphics[width=\linewidth]{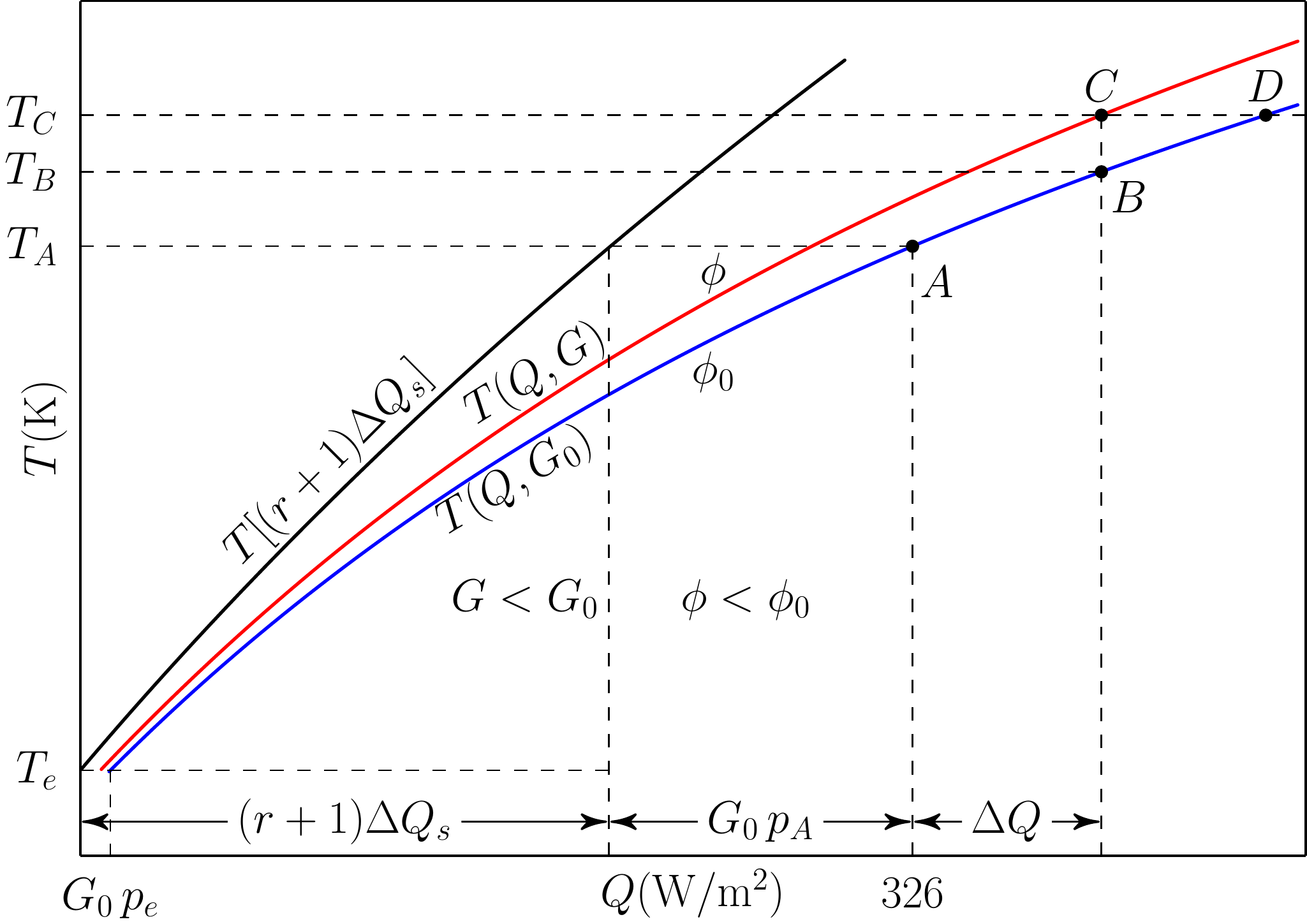}\\
  \caption{Description of the temperature change $ABC$ in $TQ$-coordinates.
  The processes $AB$ and $BC$ correspond to the first and the second term in
  Eq.~\ref{DeltaT3}, respectively.}\label{TQphi}
\end{figure}

\begin{figure}
  \centering
  \includegraphics[width=\linewidth]{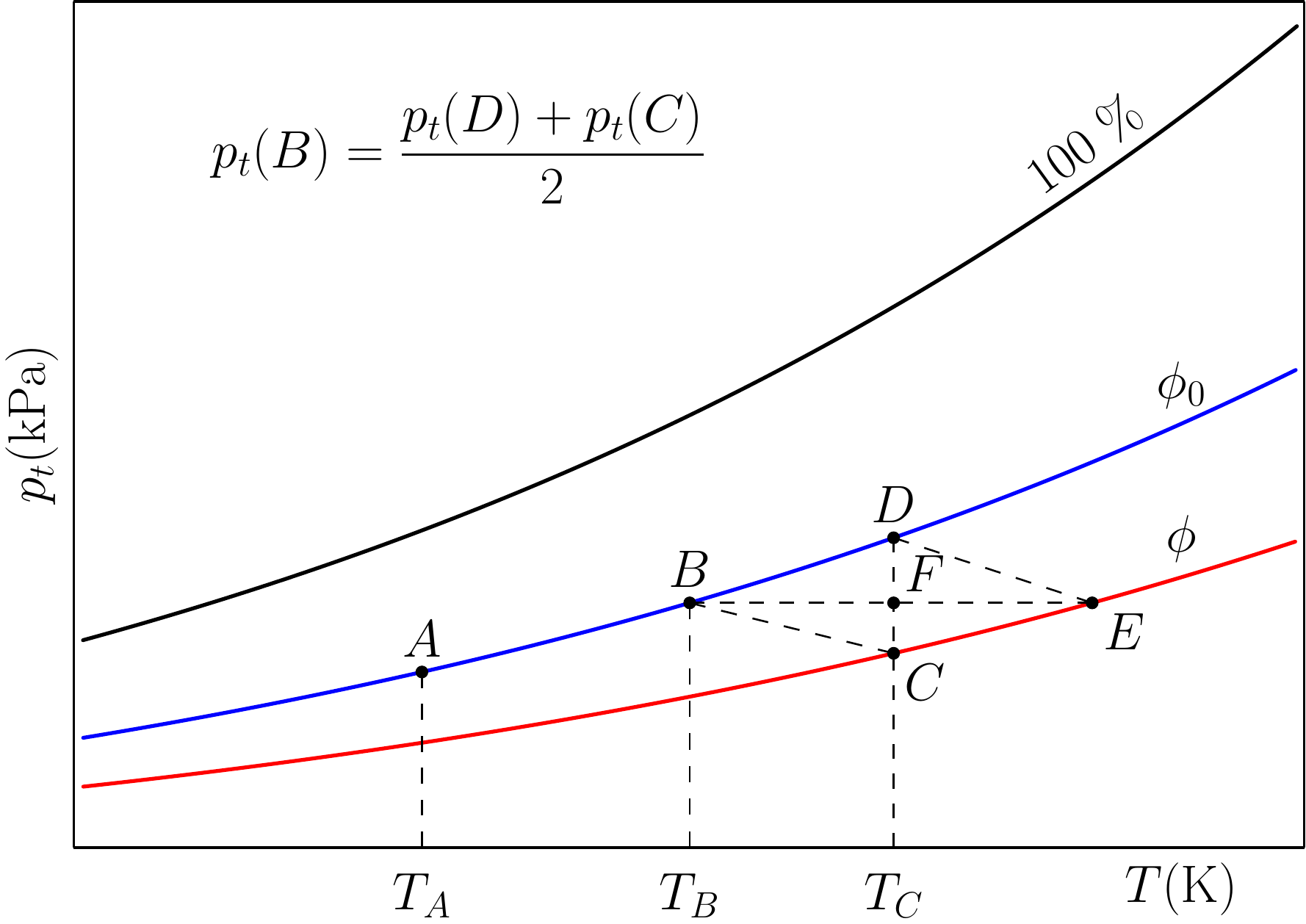}\\
  \caption{The same temperature change $ABC$ as in Fig.~\ref{TQphi} in $p_tT$-coordinates.}\label{pT}
\end{figure}

\section{Alternative New Approach for Climate Change}

As presented earlier the global mean temperature $T$ depends on the forcing $Q$ and relative humidity $\phi$ or the coefficient $G$, which is proportional to $\phi$. This means that temperature change consists of two independent terms as in Eq. (\ref{DeltaT3}). Now we will derive an alternative expression using relative humidity. The relative change of $\phi=p_t/p$ is given by
\begin{equation}\label{dphi/phi2}
  \frac{d\phi}{\phi}=\frac{dp_t}{p_t}-\frac{dp}{p},
\end{equation}
where  $dp/p=(dp/pdT)dT=\alpha dT$. Thus, $\alpha$ is the derivative of the saturated water vapour pressure with respect to the temperature divided by $p$ and it is very constant over a large temperature range. According to Eq. (\ref{dphi/phi2}) the total temperature change
\begin{equation}\label{DeltaT4}
  \Delta T=\frac{dp}{\alpha p}=\frac{1}{\alpha}\left(\frac{\Delta p_t}{p_t}-\frac{\Delta\phi}{\phi}\right),
\end{equation}
where $p_t$  and $\phi$ are global effective values in the low altitudes. Figures \ref{TQphi} and \ref{pT} shows the process $AB$, $BC$ and $AC$ in ($T,Q$)- and ($p_t,T$)-coordinates, respectively. Note that $dT$ or $\Delta T$ is mathematically an exact differential so its value does not depend on a path between the start and end point in both coordinate systems. Globally Equation (\ref{dphi/phi2}) is valid in each altitude, if the lapse rate is constant. The first term in Eq. (\ref{DeltaT4}) describes the process $AB$, where $\Delta G=0, \Delta\phi=0$ and $\Delta Q$ is nonzero. In this process
\begin{equation}\label{DeltaT_Q}
  \Delta T_Q=\frac{1}{\alpha}\left(\frac{\Delta p_t}{p_t}\right)_{AB}=R\Delta Q.
\end{equation}
The second term in Eq. (\ref{DeltaT4}) corresponds the process $BC$, where $\Delta Q=0$ but, $\Delta \phi$ and $\Delta G$ are in turns non zero. In this process for small changes according to Eq.(\ref{dphi/phi2})
\begin{equation}\label{Deltaphi/phi}
  \frac{\Delta\phi}{\phi}=\left(\frac{\Delta p_t}{p_t}-\frac{\Delta p}{p}\right)_{BC}=
  2\left(\frac{\Delta p_t}{p_t}\right)_{BC}=-2\left(\frac{\Delta p}{p}\right)_{BC},
\end{equation}
because in Fig. \ref{pT} partial pressure $p_t$ in $B$ is the average of the pressures in $D$ and in $C$. Note that $DF=FC$ in small changes. Using the above equation we can write the second term in the process $BC$ as follows:
\begin{equation}\label{DeltaT_G}
  \Delta T_G=\frac{1}{\alpha}\left[\left(\frac{\Delta p_t}{p_t}\right)_{BC}-\frac{\Delta\phi}{\phi}\right]=
  \frac{1}{\alpha}\left(\frac{\Delta\phi}{2\phi}-\frac{\Delta\phi}{\phi}\right)=
  -\frac{1}{2\alpha}\frac{\Delta\phi}{\phi}
\end{equation}
Thus we have the total temperature change in the process $AC$ given by
\begin{equation}\label{DeltaT5}
  \Delta T=\frac{1}{\alpha}\left[\left(\frac{\Delta p_t}{p_t}\right)_{AC}-\frac{\Delta\phi}{\phi}\right]=
  \frac{1}{\alpha}\left[\left(\frac{\Delta p_t}{p_t}\right)_{AB}-\frac{\Delta\phi}{2\phi}\right]=
  R\Delta Q-R\Delta Gp.
\end{equation}
Taking into account that $\Delta G=G\Delta\phi/\phi$ we have the final equation, which corresponds Eq. (\ref{DeltaT3})
\begin{equation}\label{DeltaT6}
  \Delta T = \frac{1}{\alpha}\left[\left(\frac{\Delta p_t}{p_t}\right)_{AB}-\frac{\Delta\phi}{2\phi}\right]=
  R\Delta Q-RGp\frac{\Delta\phi}{\phi}.
\end{equation}
The second term in Eq. (\ref{DeltaT3}) was $-RG(p-p_e)\,d\phi/\phi$. Now $p_e$ is not in this term, because now the red and blue curves in Fig. \ref{TQphi} do not hit the black curve. The most important result of the above equation is that it gives a test for Eq. (\ref{DeltaT3}). This means that $RGp=1/2\alpha$ or $Gdp/dT=1/2R$. If we substitute this to Eq. (\ref{Gdp/dT2}) we get $R_{00}/R=2$ or $G(dp/dT)R_{00}=1$. The results derived in this work give $G(dp/dT)R_{00}=82\cdot 0.115\cdot0.1207=1.138$ and $R_{00}/R=0.1207/0.0565=2.136$, (the first paper 2.13). The values are little larger than 1 and 2, because positive feedback is still missing in these values. Earlier we pointed out that the positive feedback can be easily taken into account by substituting $Gdp/dT$  by $Gdp/dT-1$~W/(m$^2$K). However, a correction term is not very well known from the experiments, but it is less than $-2$~W/(m$^2$K) \cite{Col,Kie1997}. If we rewrite Eq. (\ref{Gdp/dT2}) with the positive feedback $1.15~{\rm W/(m^2 K)}$, we have
\begin{equation}\label{R_tot}
  \frac{1}{R_{\rm tot}}=\frac{1}{R_{00}}+G\frac{dp}{dT}-1.15~\frac{\rm W}{\rm m^2K}=\frac{1}{R_{00}}+G_{\rm tot}\frac{dp}{dT},
\end{equation}
where $G_{\rm tot}=71.9$~W/(m$^2$kPa) and $R_{\rm tot}=0.0605$~K/(W/m$^2$K). These values give $R_{00}/R=2.0005$ and $G(dp/dT)R_{00}=1.0005$. Thus the final temperature change is given by
\begin{equation}\label{DeltaT6}
  \Delta T = R_{\rm tot} \Delta Q - R_{\rm tot} G_{\rm tot} p \frac{\Delta\phi}{\phi} = R_{\rm tot} \Delta Q - \frac{1}{2 \alpha} \frac{\Delta\phi}{\phi},
\end{equation}
where $1/2\alpha=7.8$~K. The water mass flow of this hydrological cycle is the key process in the climate. A small change in the number of condensation nuclei in unit volume changes the mass flow of the hydrological cycle resulting in changes of $G$ and $R$ and finally the change of the temperature. The water condensation depends probably on the activity of the sun, which modulates the flux of the cosmic rays. So,
\begin{equation}\label{dG/G}
  \frac{dG}{G}=\frac{dm_{\rm H_2O}}{m_{\rm H_2O}}=\frac{dw}{w}=\frac{dL}{L},
\end{equation}
where $m_{\rm H_2O}$ is the water mass flow ((kg/s)/m$^2$) and $w$ is the precipitation ((kg/s)/m$^2$). The conservation of water in the hydrological cycle gives $dG/G=dL/L$ or $G$ is dependent on $L$ which is proportional to $m_{\rm H_2O}$. So we can derive $G$ using $Z=(dp/pdT)L$ instead of the complicated expression in Eq. (\ref{Z2}). Substitution to Eq. (\ref{Gdp/dT}) gives $G=(r+1)L/p=1.507\cdot 80~{\rm W/m^2}/1.793~{\rm kPa}=67.2~{\rm (W/kPa)/m^2}$. However, equation (\ref{Z2}) gives probably a more accurate value for Z because all the eight parameters in it are measurable ones. Observe that $G_{\rm tot}$ was $71.9~{\rm (W/m^2 kPa)}$. In addition, the energy fluxes like $L$ are less accurate on the surface than on the top of the climate.

Using our theory we are able to calculate the change of $G$ and $R$ due to the change of the low cloud cover and the relative humidity. These changes indicate changes of the water mass flow. Most of water is condensating at the altitude of low cloud cover, see Fig. \ref{pilvet}. That is why we use the observed changes of the low cloud cover and the corresponding changes of the relative humidity at 700~mbar and 850~mbar. The result is that a 1~\% change in the low cloud cover changes the temperature by $-0.11^\circ$C. This result is in very good agreement with the paper by S.H. Schneider \cite{Sch1972}.

\begin{figure}
  \centering
  \includegraphics[width=\linewidth]{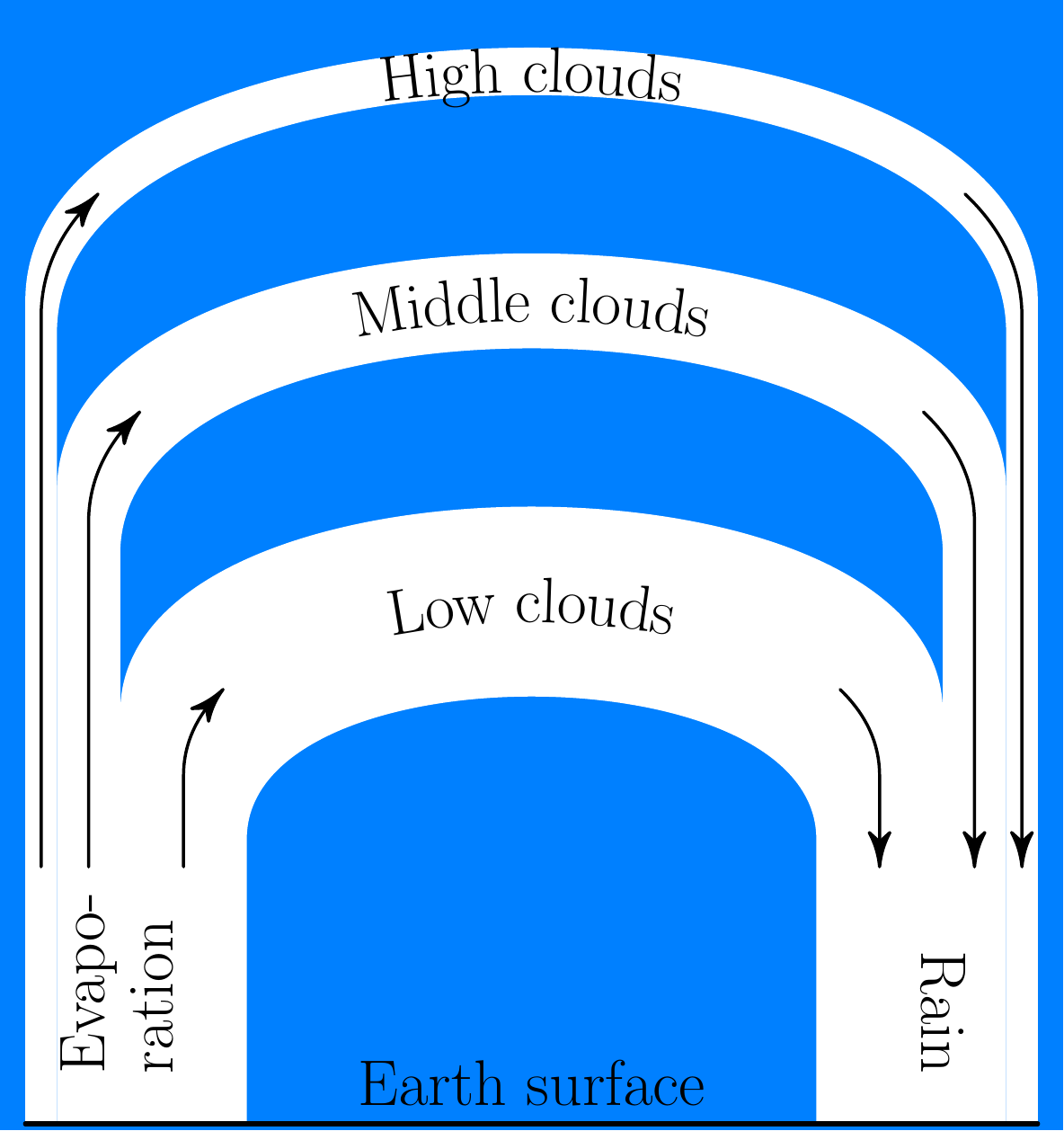}\\
  \caption{Schematic illustration of water mass flow from the earth surface
  to the clouds and back to the surface.}\label{pilvet}
\end{figure}

Because $G=G_0\phi$, where $G_0$ is a constant, we have $\Delta G=G_0\Delta\phi$ or $\Delta G/G=\Delta\phi/\phi$. Note that in Fig. 4 the feedback $G_{\rm tot}p=G_0p_t$ or is proportional to the absolute water amount in the atmosphere. In our second paper \cite{Kau2014} calculating the curve in Fig. \ref{Fig8} we used the second term $-RG(p-p_e)\Delta\phi/\phi\approx -15^{\circ}{\rm C}\Delta\phi$. The corresponding theoretical value was $-RG(p-p_e)\Delta\phi/\phi\approx -17^{\circ}{\rm C}\Delta\phi$. The new theory gives
\begin{equation}\label{RGp}
  -R_{\rm tot}G_{\rm tot}p = -\frac{1}{2\alpha} \frac{\Delta\phi}{\phi} = -7.8^{\circ}{\rm C} \frac{\Delta\phi}{\phi} = -14.2^{\circ}{\rm C}\Delta\phi.
\end{equation}
The global warming has almost stopped about 16 years ago and the temperature has slightly decreased between years 2002 and 2012. The explanation for this behavior is simply the turning point in the relative humidity around year 2002. This is clearly shown in our second paper \cite{Kau2014}. Look at 700~mbar curve in Fig 2.

\section{Conclusion and Discussion}

IPCC has used in their estimations for example the results of the paper ``Thermal Equilibrium of the Atmosphere with a Given Distribution of Relative Humidity'' by Syukuro Manabe and Richard T. Wetherald \cite{Man}. The authors have used in their calculations a heat capacity of air as a heat capacity of the whole atmosphere. The heat capacity of the mere air is about $10~{\rm MJ/m^2K}$. However, in a correct calculation we have to use the capacities $10.8~{\rm MJ/m^2K}$ over land and $325~{\rm MJ/m^2K}$ over ocean. They have not added the heat capacity of a thin layer of the ground over land and the mixing layer (75~m) of the ocean. Taking into account the fact that the earth consists of 29~\% land and 71~\% ocean we can estimate the effective heat capacity of the whole climate. It is about $60~{\rm MJ/m^2K}$ or six times bigger than the value used by Manabe et al. They derived the response time roughly between 30 and 60 days, which are in a good agreement with the observations. See Fig.~6 in their paper \cite{Man}.

According to Physics the response time is the product of sensitivity and capacity. However, we cannot only multiply these numbers by six, because it gives too long response time, between 180 and 360 days. So we have also to divide the sensitivity by six so that the product of the sensitivity and the heat capacity is the response time between 30 and 60 days. Because they have used only the sixth part of the real heat capacity they obtained six times bigger sensitivity. In their paper the calculated sensitivities for doubling CO$_2$ were $2.3^{\circ}$C and $1.3^{\circ}$C for the atmosphere with the realistic distribution of relative humidity and with the realistic distribution of absolute humidity, respectively. So the more realistic sensitivities are those numbers divided by 6 or $0.38^\circ$C and $0.27^{\circ}$C. These sensitivities are very comparable with our result $0.24^{\circ}$C \cite{Kau2014}. The same mistake is repeated later, at last in the year 1975.

If the climate sensitivity were the IPCC value $\Delta T_{\rm 2CO_2} \approx 3.2^\circ$C, the warmest time of the year would be around September 15 in the northern hemisphere, but according to the observations it is around July 25. This is a strong proof against the results of the circulation models.

We have derived the climate sensitivity at the present climate condition using solely the observed energy budget of the climate. In our first paper \cite{Kau2011} we had only one observed point $B$ and the estimated point $A$ (See Fig. 4). We assumed that the negative feedback is proportional to the amount of water in the climate, roughly $Gp=G_0p_t$. In this work we have used 12 observed values of the climate from the energy budget. However, these values give the sensitivity 0.0565~K/(W/m$^2$), which is very close to the sensitivity $0.058~{\rm K/(W/m^2)}$, the result of our paper \cite{Kau2011}. Note, that the sensitivity $R_{\rm tot} = 0.0605~{\rm K/(W/m^2)}$ includes the small positive feedback. The main difference between these studies is in the sensitivity values without feedback. In the earlier paper \cite{Kau2014} $R_0=dT/dQ_s = 0.183~{\rm K/(W/m^2)}$ because we set $dQ_s=dQ$ in long wave IR emission. In the present paper we used the energy budget and we were able to use the relation $dQ_e=rdQ_r$, ($r=0.507$). This gives $(1+r)dQ_s=dQ$ or $R_{00}=R_0/(1+r)=0.121$~K/(W/m$^2$). However, the value of $G=82$~(W/m$^2$)/kPa compensates this difference. Note that $G$ was 103~(W/m$^2$)/kPa) in our previous paper \cite{Kau2011}. As a final conclusion the $T(Q)$-curves are almost identical in both studies.

In this work we have neglected all the small effects, which change the sensitivity one percent or less, for example a change of the surface albedo, warming of water in rainfall, a small change of the constant $r$. Of course, it is possible to take into account small effects mentioned above, but we cannot verify the results experimentally. The reason is that uncertainties in the observed data of the global mean temperature, the low cloud cover, and the relative humidity are still too large.

\bibliographystyle{amsplain}
\bibliography{climate}
\end{document}